\documentclass[12pt,preprint]{aastex}
\usepackage{graphicx}
\usepackage{amssymb,amsmath}

\newcommand\beq{\begin{equation}}
\newcommand\eeq{\end{equation}}
\newcommand\lsim{\mathrel{\rlap{\lower4pt\hbox{\hskip1pt$\sim$}}
        \raise1pt\hbox{$<$}}}
\newcommand\gsim{\mathrel{\rlap{\lower4pt\hbox{\hskip1pt$\sim$}}
        \raise1pt\hbox{$>$}}}

\begin{document}
\title{Atmospheres of Hot Super-Earths}

\author{Thibaut Castan,\altaffilmark{1,2} and Kristen
  Menou\altaffilmark{2}}
 
\altaffiltext{1}{Departement de Physique,
Ecole Normale Superieure, 24 rue Lhomond, 75005 Paris, France}
\altaffiltext{2}{Department of Astronomy, Columbia University, 550
  West 120th Street, New York, NY 10027}

\begin{abstract}
Hot super-Earths likely possess minimal atmospheres established
through vapor saturation equilibrium with the ground. We solve the
hydrodynamics of these tenuous atmospheres at the surface of Corot-7b,
Kepler 10b and 55 Cnc-e, including idealized treatments of magnetic
drag and ohmic dissipation. We find that atmospheric pressures remain
close to their local saturation values in all cases. Despite the
emergence of strongly supersonic winds which carry sublimating mass
away from the substellar point, the atmospheres do not extend much
beyond the day-night terminators.  Ground temperatures, which
determine the planetary thermal (infrared) signature, are largely
unaffected by exchanges with the atmosphere and thus follow the
effective irradiation pattern. Atmospheric temperatures, however,
which control cloud condensation and thus albedo properties, can
deviate substantially from the irradiation pattern. Magnetic drag and
ohmic dissipation can also strongly impact the atmospheric behavior,
depending on atmospheric composition and the planetary magnetic field
strength. We conclude that hot super-Earths could exhibit interesting
signatures in reflection (and possibly in emission) which would trace
a combination of their ground, atmospheric and magnetic properties.
\end{abstract}

\section{Introduction}

Hot super-Earths are an emerging class of exoplanets. CoRot 7b, Kepler
10b and 55 Cnc-e are currently the only well-characterized, transiting
members of this new class (Leger et al. 2009; Batalha et al. 2011;
Fischer et al. 2008; Winn et al. 2011; Demory et al. 2011) but
observational trends suggest that many more hot super-Earths will be
discovered and characterized in the future (e.g., Howard et al. 2011;
Borucki et al. 2011).

Structural and evolutionary considerations suggest that hot
super-Earths are rocky planets which have lost their original volatile
atmospheres, and possibly even some of their rocky material, via
atmospheric erosion (Valencia et al. 2010). Indeed, a hot rocky
super-Earth will retain a minimal atmosphere that is continuously
replenished via vapor saturation equilibrium with the ground (Schaefer
\& Fegley 2009; Leger et al. 2011) and is thus perpetually subject to
erosion. Hot super-Earths are also expected to be tidally-locked to
their parent stars. Under these conditions, large temperature and
surface pressure differences will exist between the day and the night
sides of these planets. By contrast with the majority of known
planetary atmospheres, which have a well distributed atmospheric mass
around the planet, an unusual circulation regime will thus develop at
the surface of hot super-Earths, with powerful winds dynamically
redistributing atmospheric mass via exchanges with the ground.

Understanding the atmospheric behavior of hot super-Earths is
important for the interpretation of direct observational data and
because it informs evolutionary considerations such as atmospheric
erosion.  In this Letter, we study the unusual hydrodynamics of the
atmosphere of hot super-Earths, using a simple model inspired by
related work on Io (Ingersoll et al. 1985). In \S2, we describe our
model hypotheses, equations and method of solution. Our main results
are presented in \S3. We conclude by discussing some potential
implications of our work in \S4.

\section{Model}

\subsection{Hypotheses}

We consider hot super-Earths that are tidally-locked to their parent
stars and thus possess permanent day and night sides. We treat the
atmosphere as a continuous fluid that is hydrostatically bound to the
planet. The fluid approximation is justified, even if only marginally
so in the most tenuous regions described by our solutions, where one
runs out of atmosphere near the planetary day-night terminator. For
simplicity, we ignore rotation and Coriolis effects in our analysis,
so that point symmetry around the substellar point can be
assumed. Comparing the rotation timescale ($1/\Omega$) for the three
super-Earths of interest with the advection time across a planetary
radius ($r/V$) we find Rossby numbers $\sim 2$ for the typical
velocities $\sim$~km/s obtained in our solutions. It is thus
reasonable to neglect rotation as a first approximation but improved
models may need to account for it.\footnote{Rotation could break the
  substellar-point symmetry of our solutions and, under conditions of
  permanent hemispheric forcing, could lead to the formation of
  superrotating equatorial winds (Showman \& Polvani 2011).}

The atmospheric composition that results from vapor saturation
equilibrium with the ground on a hot super-Earth is a priori
unknown. The analysis of Schaefer \& Fegley (2009) for such sublimating
atmospheres suggests that monatomic sodium is the dominant constituent
on a hot super-Earth with a bulk silicate earth composition, prior to
any fractional loss. According to their analysis, this is also the
case with the largest overall atmospheric mass (surface
pressures). Therefore, without loss of generality, we adopt a pure
sodium composition in our models, so as to maximize atmospheric
effects in the solutions, and we comment on the consequences of
adopting other compositions. Since the atmospheres of interest are
tenuous, typically $100$ mbar or much less, we assume that the
absorption optical depths in the thermal and the visible are
negligibly small ($\ll 1$). Non-radiative considerations, in
particular dynamics and exchanges with the ground, will then determine
the thermodynamical state of the atmosphere.

We postulate, and verify a posteriori, that latent heat exchanges with
the atmosphere have at most a small effect on the surface energy
budget, which is dominated by radiative fluxes. We therefore assume
that the ground temperature, $T_s$, can be obtained by simple
radiative balance for the permanent dayside: $T_{s}=
(T_{sub}-T_{as})\cos^{1/4}\theta+T_{as}$ for $\theta <85\char23$,
where $\theta$ is the angle away from the substellar point, $T_{sub}$
is the substellar temperature, and $T_{as}$ is the antistellar
temperature. For $85\char23 \le \theta < 110\char23$, we assume that
the temperature decreases linearly with $\theta$ to $T_{as}=50$~K, a
value which may be reasonable on the basis of geothermal heating of
the nightside (Leger et al. 2011), with little consequences on our
results.  The substellar surface temperature is evaluated simply as
$T_{sub}= T_\star \sqrt{R_{\star}/D}$, where $T_\star $ and $R_\star$
are the stellar effective temperature and radius, $D$ is the
planet-star orbital separation and a negligible albedo is assumed.

\subsection{Equations}

Our formalism is directly inspired from that used by Ingersoll et
al. (1985) for the study of a frost atmosphere on Io. Indeed, we
validated our implementation by reproducing the results for Io first,
before moving on to super-Earths. Like Ingersoll et al. (1985), we
consider that strong vertical exchanges in the thin, turbulent
atmosphere justify the use of vertically-integrated equations for the
conservation of atmospheric mass, momentum and energy. Assuming that
the velocity and the entropy per unit mass (for a dry adiabat) are
constant with height in the atmosphere,\footnote{Different assumptions
  for these vertical profiles would lead to qualitatively similar
  results, as shown explicitly by Ingersoll et al. (1985) with a dry
  vs. wet adiabat comparison. We note, however, that these assumptions
  could become invalid if sufficient stellar energy is deposited in
  the atmosphere for a stably stratified thermal profile to be
  established.}  the governing equations for the pressure, temperature
and velocity at the base of the atmosphere can be written (see
Ingersoll et al. 1985 for details):

\beq \label{eq:1}
\frac{1}{rg\ \sin\theta} \frac{d}{d\theta}\left(VP\ \sin\theta\right)
\ =\ mE,
\eeq
\beq \label{eq:2}
\frac{1}{rg\ \sin\theta} \frac{d}{d\theta}\left[\left(V^{2}\ +\ \beta
C_{p}T \right)\ P\ \sin\theta \right]\ =\ \frac{1}{rg\ \tan\theta}\beta
 C_{p}TP \ +\ \tau,
\eeq
\beq \label{eq:3}
\frac{1}{rg\ \sin\theta} \frac{d}{d\theta}\left[ \left( \frac{V^{2}}{2}
\ +\ C_{p}T \right)VP\ \sin\theta \right]\ =\ Q,
\eeq

where $r$ is the planetary radius and $g$ is the planet's surface
gravity. The three unknowns of the atmospheric flow are the velocity,
$V$, the pressure, $P$, and the temperature, $T$, at the base of the
atmosphere. The thermodynamic parameter $ \beta ={\cal
  R/(R}\ +\ C_{p}) $, where ${\cal R}=k_{\rm B}/m$ is the gas constant
and $C_{p}$ is the specific heat at constant pressure. For sodium,
with a mean molecular weight $\mu=23$ and a mass per atom $m = \mu
m_{\rm H}$, we adopt the monatomic values
$C_{p}=904$~J~K$^{-1}$~kg$^{-1}$ and $\beta=0.286$.

$E$, $\tau$, and $Q$ are the rates per unit area of molecules,
momentum and energy transferred to the atmosphere by exchange with the
ground. These fluxes, which are mediated by a boundary layer of
negligible vertical extent at the base of the atmosphere, are governed
by sublimation/condensation processes, flow advection, and turbulent
exchanges by eddies. Our implementation of these surface-atmosphere
boundary layer exchanges follows closely that adopted by Ingersoll et
al. (1985). The net flux of sublimating/condensating molecules is
proportional to the difference between the vapor pressure $P_{v}$ of
the surface and the atmospheric pressure $P$, so that
$E\ =\ {(P_{s}\ -\ P)}/({v_{s}\sqrt{2\pi}})$, where $v_{s}=
(kT_{s}/m)^{1/2}$ is the molecular thermal speed at the surface. We
approximate the vapor pressure curve of sodium by fitting the sodium
curve for the bulk silicate atmosphere model shown in Figure 1 of
Schaefer \& Fegley (2009) with the formula $P_{s}(T_s)=A
\exp(-B/T_s)$. We adopt $A=10^{9.6}$~Pa and $B=38,000$~K.

The fluxes $Q$ and $\tau$ depend linearly on the quantities being
transported, with the same transfer coefficients $w_{s}$ and
$w_{a}$. Energy gains from the surface are proportional to the
enthalpy per unit of mass at the surface, $C_{p}T_{s}$, while losses
scale as the sum of the kinetic energy and enthalpy per unit mass,
$(V^{2}/2\ +\ C_{p}T)$. For momentum, gains from the surface are zero,
while losses are proportional to $V$. This leads to the formulation

\beq
\tau\ =\ -\rho_{s}w_{a}V,
\eeq
\beq
Q\ = \ \rho_{s} w_{s}C_{p}T_{s} \ -\ \rho_{s}w_{a}(V^{2}/2\ +\ C_{p}T),
\eeq

where $\rho_{s}= m P_{s}/(k_{b}T_{s})$ is the density in the boundary
layer.  Two different contributions are included in the definition of
the transfer coefficients $w_{a}$ and $w_{d}$. First, the contribution
from the mean flow normal to the surface is proportional to $E$ and is
represented by the velocity $ V_{e}=mE/\rho_{s} $. Second, the
contribution from turbulent eddies is obtained from turbulent boundary
layer results in the case $E=0$. We adopt the same expressions for the
transfer coefficients $w_{a}$ and $w_{d}$ as Ingersoll et
al. (1985). Finally, where needed, we use Sutherland's ideal gas
formula to evaluate the dynamic viscosity of the atmospheric gas:
$\eta\ =\ \eta_{0}({T_{s}}/{T_{0}})^{3/2}({T_{0}+C})/({T_{s}+C})$,
with $\eta_{0}=1.8 \times 10^{-5}$~kg~m$^{-1}$~s$^{-1}$, $T_{0}=291$~K
and $C=120$~K. We have verified that our results do not strongly
depend on details of the viscosity law adopted.

In sufficiently ionized regions, magnetic drag and ohmic dissipation
can influence atmospheric flows on hot super-Earths, if these planets
possess strong enough magnetic fields (Gaidos et al. 2010; Tachinami
et al. 2011; Driscoll \& Olson 2011). Here, we build on the recent
understanding gained in the context of hot giant exoplanet atmospheres
(Batygin \& Stevenson 2010; Perna et al. 2010a,b; Menou
2011). Assuming induced electric currents that are confined to the
thin atmosphere of hot super-Earths, magnetic drag and ohmic
dissipation may be included in the simplest possible way as
additional, vertically-integrated contributions to the fluxes of
momentum and energy received by the atmosphere, through linear terms
proportional to the inverse of a representative magnetic drag time,
$T_{drag}$. Therefore, in models including magnetic effects, we add
contributions to $\tau$ and $Q$ which act to brake the winds and heat
up the atmosphere according to:

\beq
\tau_{\rm mag}\ =\ - \frac{PV}{g T_{drag}},~~~  Q_{\rm mag}\ =+\frac{PV^{2}}{g T_{drag}}.
\eeq

Adopting a magnetic field strength $B=1$~G, detailed estimates of the
resistivity ($\eta$) based on the formalism described in Menou (2011)
yield drag times $T_{drag} ~\propto \eta/B^2$ that vary greatly over
the planet for a sodium atmosphere. Drag times can be much shorter
than a typical advection time, $r/V$, in the vicinity of the hot
substellar point, but they can become negligibly large in
poorly-ionized regions where $T \lsim 1000$~K. Changing the assumed
composition to a less easily ionized atom, e.g. from Na to Mg (another
possibly abundant constituent; Schaefer \& Fegley 2009), can also lead
to dramatically reduced magnetic effects. Therefore, rather than
exploring magnetic effects in detail here, we illustrate their
consequences below by presenting two idealized models for Kepler 10-b
with spatially uniform values of $T_{drag} = 10^4$ and $5 \times
10^3$~s, which have been chosen to match the typical advection time in
our solutions with velocities $\sim$ km/s.

\subsection{Method of Solution}

We solve our set of differential and auxiliary equations for the
unknowns $P,\ T,\ V$ with a relaxation method (specifically, the
algorithm described in Hameury et al. 1998). Special care must be
taken when integrating these equations because the solution goes
through a sonic point at an unspecified angle, $\theta_{s}$. To find
$\theta_{s}$, we define $\theta$ as a fourth variable on a numerical
domain describing the subsonic region only and we apply physically
sensible boundary conditions on this domain.

At the substellar point, $V(\theta =0)=0$ is imposed for symmetry and
$T(\theta=0)=T_{sub}$. The transonic solution must have a well-behaved
value of $dV/d\theta$ at the sonic point, which provides the two
additional boundary conditions needed on the subsonic domain.  Noting
that the following first-order differential equation is satisfied by the
velocity,

\beq \label{eq:dvdt}
Cp P \left( T -\frac{V^{2}}{{\cal R}} \right)\frac{dV}{d\theta}=\left(\frac{1}{\beta}-\frac{1}{2}\right)mgrV^{2}E-\frac{rgV\tau}{\beta}+rgQ-\frac{C_{p}TPV}{\tan \theta},
\eeq

requiring a smooth sonic transition leads to the two boundary
conditions at $\theta=\theta_{s}$:

\begin{equation*}
V=\sqrt{{\cal R}T},\ \ \ P=gr\tan\theta \times \frac{{\cal R}}{C_{p}}\left(\frac{Q}{V^{3}}-\frac{\tau}{\beta V^{2}}+\left(\frac{1}{\beta}-\frac{1}{2}\right)\frac{mE}{V}\right).
\end{equation*}

Our initial guesses for the relaxation method are based on the
analytical solutions of Ingersoll (1989), adapted to our super-Earth
problem. Once $\theta_{s}$ and the solution on the subsonic domain are
known, using a small extrapolation across the sonic point, the
integration can proceed in the supersonic domain using a simple Euler
method. The integration is stopped when the pressure is so low that
the fluid approximation ceases to be valid.

\section{Results}

We solve the hydrodynamics of sodium atmospheres at the surface of
CoRot-7b, Kepler 10b and 55 Cnc-e (see Table 1 for the specific
parameters adopted)\footnote{Although 55 Cnc-e may require a
  surprisingly low-density, Moon-like bulk composition to be
  considered purely rocky (Demory et al. 2011), we include it in our
  analysis as a useful probe of the super-Earth parameter
  space.}. Figures 1-4 show the pressure, temperature, velocity and
Mach number obtained for the base of the atmosphere, as a function of
the angular distance from the substellar point. In each figure, large
dots indicate the location of the sonic points. Solutions are shown as
red long-dashed lines for CoRot-7b, black dotted lines for Kepler 10b
and blue solid lines for 55 Cnc-e. In addition, two solutions
illustrating magnetic effects for Kepler 10-b are shown as a yellow
dash-dotted line ($T_{drag}= 10^4$~s) and a green short-dashed line
($T_{drag}= 5 \times 10^3$~s). Finally, a thick black solid line in
Figure 2 shows the ground temperature in the Kepler 10-b model, for
comparison with the atmospheric temperatures.

Ignoring magnetic effects, our atmospheric solutions for the three hot
super-Earths share many similarities. CoRot 7-b is essentially a
cooler, more tenuous version of Kepler 10b and 55 Cnc-e. This
similarity stems from the strong control that vapor saturation
equilibrium exerts on the atmospheric flow, via the exponential
dependence of pressure on surface temperature, with a comparatively
small influence for other model parameters such as planetary radius or
surface gravity.

In these atmospheric flows, winds are accelerated by conversion of
thermal and gravitational potential energy into kinetic
energy. Supersonic speeds are reached at $\theta_s \sim
40$~deg. Atmospheric pressures remain close to the local saturation
values, within tens of \% or so, over most of the flow, with stronger
deviations in the end region at $\theta \gsim 80$~deg. Distinct
regions of atmospheric sublimation ($E>0$) and condensation ($E<0$)
roughly correspond to the subsonic and the supersonic regions of the
flow, respectively. Beyond $\theta \sim 85$-$90$~deg, where the
atmosphere becomes very tenuous, viscous drag becomes very strong and
the flow quickly transitions into a regime where kinetic energy is
returned into thermal and gravitational energy. Large Mach numbers are
achieved before the atmospheric flow effectively comes to a
halt. These flow properties are generally consistent with the ones
described by Ingersoll et al. (1985) for a sulfur dioxyde frost
atmosphere on Io, despite the vastly different thermodynamic regime
realized on hot super-Earths. Therefore, we expect these results to
hold rather generally for hot super-Earths, including for different
atmospheric compositions, even if a reduced angular extent of the flow
(limited by strong viscous drag) can be expected for less massive
atmospheres.

Magnetic effects complicate this picture significantly. Magnetic drag
slows down the winds in the subsonic region, as expected, but larger
wind speeds are ultimately reached in the supersonic region. This
delayed acceleration can be understood as resulting from the
additional thermal and gravitational energy made available to the flow
by significant ohmic heating, especially around and beyond the sonic
point (compare the temperature profiles of models with and without
magnetic effects in Fig.~2). The atmospheric solutions with uniform
values of $T_{drag} = 10^4$ and $5 \times 10^3$~s for Kepler-10b
differ significantly from each other and the unmagnetized case,
showing that magnetic effects can have a strong impact on the
flow. One should remember that our treatment of these effects remains
greatly idealized: magnetic drag may preferentially act on the zonal
component of the flow for a dipolar planetary field and the magnitude
of this drag (and ohmic dissipation) will vary exponentially with
atmospheric temperature (e.g., Menou 2011). The large range of
temperatures realized on hot super-Earths suggests a strongly magnetic
atmospheric flow near the substellar point and a largely unmagnetized
flow far from it (where $T \lsim 1000$~K). These important
considerations are not addressed by our models.

Our solutions justify a posteriori the use of a fixed surface
temperature profile, controlled by radiative fluxes. The evaporation
rate peaks at a few $10^{24}$~m$^{-2}$~s$^{-1}$ at the substellar
point, drops to zero near the sonic point and reaches minimal values
of minus a few $10^{23}$~m$^{-2}$~s$^{-1}$ around $\theta \sim
60$~deg. Given a latent heat of sublimation ${\cal{L}_{\rm sub}} \sim
10^5$-$10^6$~J~kg$^{-1}$ for atmospheric compositions of interest
(e.g., Valencia et al. 2010), we estimate deviations from the
radiative temperature profile due to latent heat effects which are at
most $\sim$ a few \% at the substellar point. They can be neglected in
view of other sources of uncertainties in the model.

\section{Discussion and Conclusion}

According to our solutions, the range of atmospheric temperatures and
pressures realized on the day-side of a hot super-Earth is
vast. Rather than a pure composition like sodium, it is likely that
such an atmosphere would also include minor constituents which are
subject to condensation into clouds, as discussed by Schaefer \&
Fegley (2009). The diverse range of thermodynamic conditions in the
atmosphere across the dayside opens the possibility for the formation
of many different classes of clouds. If some of these clouds are thick
and reflective enough, they could dominate the planetary albedo
properties of a hot super-Earth, given surface albedos of only $\sim
0.1$ (Leger et al. 2011). This may lead to interesting signatures in
the reflected light from a hot super-Earth, if for instance strong
albedo variations occur across the planetary dayside. A large enough
albedo from clouds ($\gsim 0.1$) would also generate variations in the
surface irradiation pattern, which would imprint an additional
signature in the dayside thermal emission of the planet. Excitingly,
the measurable optical phase curve of Kepler 10b could eventually
reveal such signatures (Batalha et al. 2011).

Our results show that, even for a relatively massive sodium
atmosphere, the powerful winds which carry sublimating mass away from
the substellar point on a hot super-Earth are unable to establish a
significant atmosphere beyond the planetary day-night terminator. This
conclusion impacts atmospheric erosion scenarios. Traditional
assumptions of spherical symmetry and of an atmosphere at rest are
poorly justified for hot super-Earths. While magnetized wind scenarios
favor atmospheric loss from the polar regions (e.g., Adams 2011;
Trammel et al. 2011), where there can be little atmosphere
according to our models, it can also be the case that atmospheric
temperatures in the polar regions are low enough that magnetic
coupling can be neglected. Furthermore, at surface pressures $\lsim
10^{-2}$-$10^{-4}$~Pa, which are reached near the terminator in our
solutions, the most tenuous part of the atmosphere may become
transparent to UV irradiation (Valencia et al. 2010), which would
presumably stall erosion.
 
We conclude by noting that the sublimation/condensation rates obtained
in our solutions suggest the possibility of substantial planetary
resurfacing by atmospheric transport. The global atmospheric
circulation carries mass from the substellar region to a region
centered around $\theta \sim 50$-$70$~deg, where net condensation
occurs. Over Gyr timescales, this process could impact the atmospheric
composition by preferentially transporting some (sublimating)
constituents relative to others and, ultimately, coupling the
availability of volatiles at the substellar point to internal
geodynamic redistribution processes. In principle, strong resurfacing
may also influence the tidal response of the rocky planet by gradually
modifying its gravitational moments, if condensation occurs beyond the
angular extent of the dayside magma ocean (e.g., Leger et
al. 2011). Detailed calculations will be needed to evaluate the
magnitude of these effects.

\acknowledgments

KM thanks Peter Goldreich for bringing to his attention the work of
Ingersoll et al. on Io. This work was supported in part by NASA grant
PATM NNX11AD65G.

\newpage

\begin{table}[h!]
\begin{tabular}{ | c | c | c | c | }
 \hline                       
Parameter & Kepler-10b & Corot-7b & 55 Cnc e \\
\hline
Stellar effective temperature, $T_{\star}$ (K) & 5630 & 5250 & 5370\\
Stellar radius, $R_{\star}$ (m) & $7.3 \times 10^{8}$ & $5.7 \times 10^{8}$ & $ 8.0 \times 10^{8}$ \\
Orbital distance, D (m) & $2.5 \times 10^{9}$ & $2.6 \times 10^{9}$ & $2.3 \times 10^{9}$\\
Planetary radius, r (m) & $9.0 \times 10^{6}$ & $10.1\times 10^{6}$ & $12.8 \times 10^{6}$ \\
Planetary surface gravity, g (m s$^{-2}$) & 22.2 & 27.1 & 19.6 \\
Planetary substellar temperature, $T_{sub}$ (K) & 3040 & 2470 & 2970 \\
\hline
\end{tabular}
\caption{Parameters for Hot Super-Earth Models.}
\label{tab:table}
 \end{table}

\begin{figure*}[l]
\centering \includegraphics[scale=0.8]{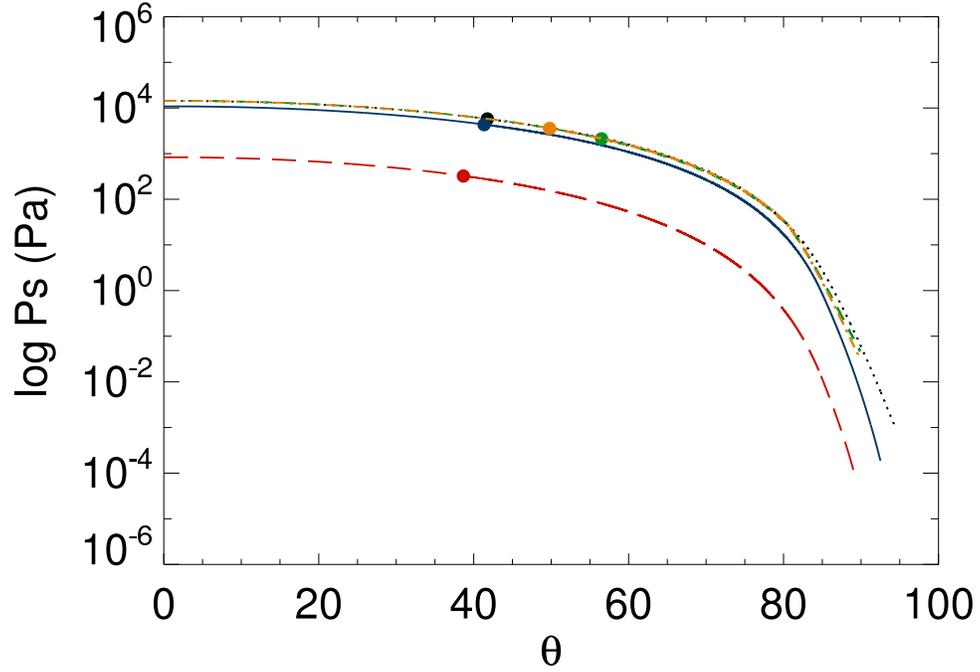}
\caption{Atmospheric pressure (in Pascal) as a function of the angular
  distance from the substellar point. Blue solid, red long-dashed and
  black dotted lines show the results for a sodium atmosphere on 55
  Cnc e, Corot-7b, 55 and Kepler-10b, respectively. Solutions with
  magnetic effects for Kepler-10b are also shown with yellow
  dash-dotted ($T_{drag}=10^4$~s) and green short-dashed ($T_{drag}=5
  \times 10^3$~s) lines.  Dots represent the sonic point location.}
\label{fig:Pressions}
\end{figure*}

\begin{figure}[h!]
\centering \includegraphics[scale=0.8]{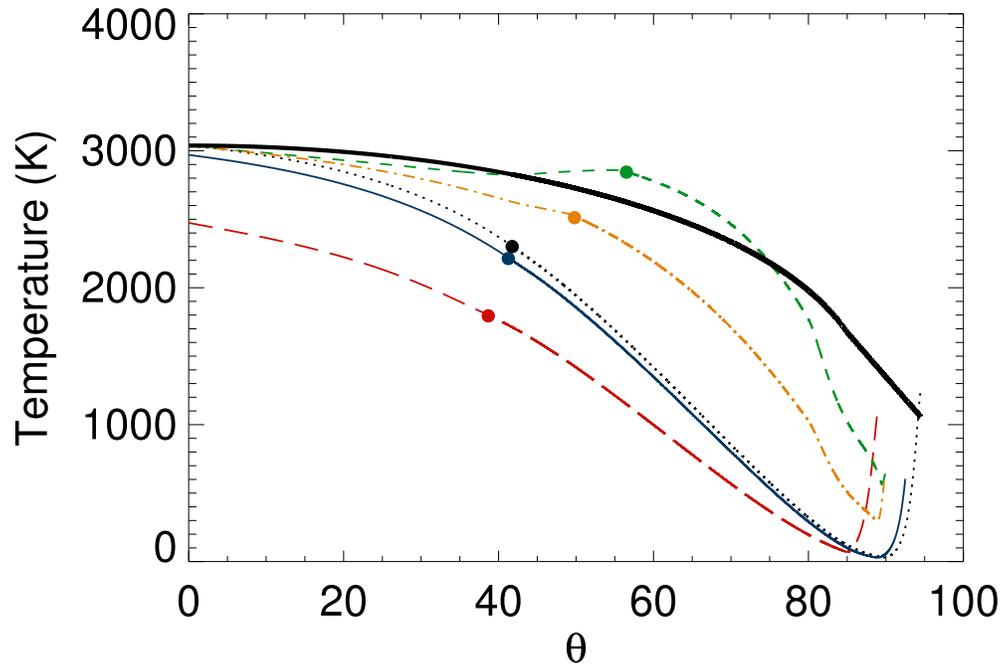}
\caption{Temperature (in K) at the base of the atmosphere as a
  function of the angular distance from the substellar point. Same
  notation as in Fig.~1. A black thick solid line also traces the surface
  temperature profile adopted in the Kepler 10b models.}
\label{fig:Temperatures}
\end{figure}

\begin{figure}
\centering \includegraphics[scale=0.8]{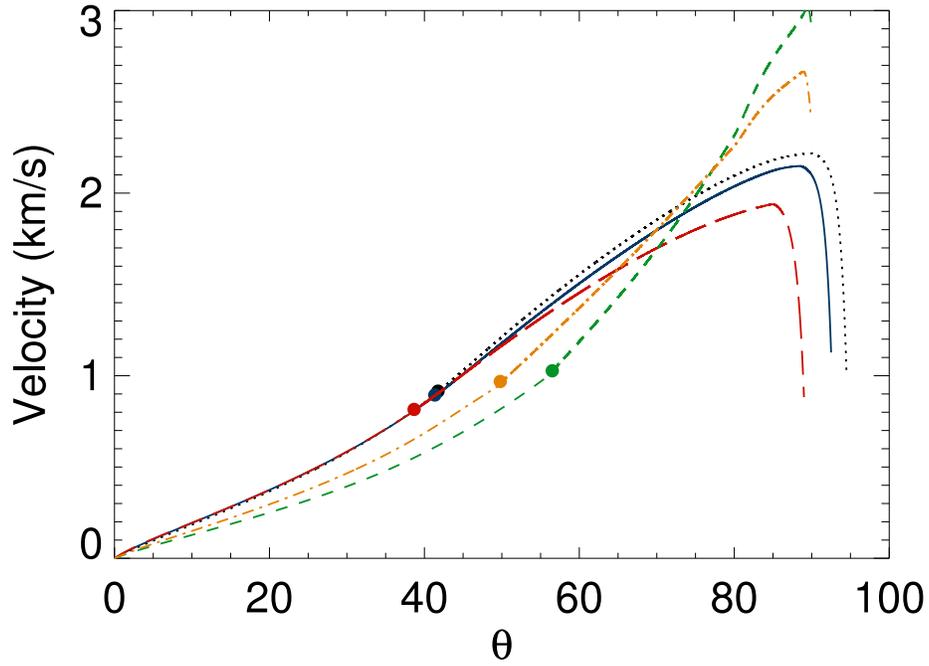}
\caption{Velocity (in km/s) at the base of the atmosphere as a
  function of the angular distance from the substellar point. Same
  notation as in Fig.~1.}
\label{fig:Vitesses}
\end{figure}

\begin{figure}
\centering \includegraphics[scale=0.8]{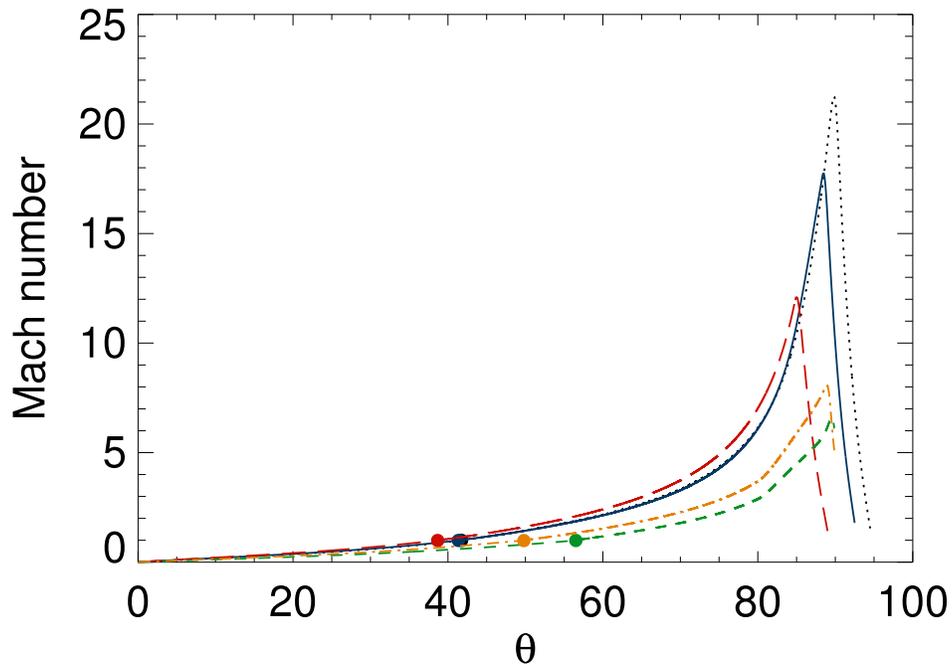}
\caption{Mach number of the atmospheric flow as a function of the
  angle from the substellar point. Same notation as in Fig.~1.}
\label{fig:Mach}
\end{figure}

\end{document}